\documentclass[twocolumn,custom,fleqn]{ModDetSymp}

\usepackage{geometry}
\usepackage[labelsep=period,labelfont=normalfont,justification=justified, singlelinecheck=false]{caption}
\usepackage{fix2col}
\usepackage{ifthen}
\usepackage{subfig}
\usepackage{graphicx}

\usepackage{hyperref}

\usepackage{cuted}

\usepackage{float}
\usepackage{stfloats}

\usepackage{bigstrut}
\usepackage{amsmath,mathtools}
\usepackage{adjustbox}
\usepackage{amssymb}
\allowdisplaybreaks

\usepackage{booktabs,array,makecell} % For professional looking tables
\usepackage{multirow}
\usepackage{dcolumn}
\usepackage{threeparttable} 
\usepackage{hhline}
\usepackage{bigstrut}

\usepackage{color,soul}

\usepackage{afterpage}

\usepackage{printlen}

\usepackage{siunitx}
\newcommand{\Ts}{\rule{0pt}{2.6ex}}       % Top strut
\newcommand{\pde}[2]{\frac{\partial #1}{\partial #2}}
\newcommand{\ode}[2]{\frac{\mathrm{d} #1}{\mathrm{d} #2}}
\newcommand{\mde}[2]{\frac{\mathrm{D} #1}{\mathrm{D} #2}}

\usepackage{calc}

\usepackage{mhchem}

\usepackage[draft]{changes}
\definechangesauthor[name={shem}, color=blue]{shem}
\graphicspath{{Figs/}}

\title{Estimating Energy Release in Metallized Composite Explosives Using the Taylor Model}

\author{Jason Loiseau$^{\dag}$, Sebastian Rodriguez Rosero$^{\ddag}$, Yaroslava Poroshyna$^{*}$, S. She-Ming Lau-Chapdelaine$^{\dag}$}

\affiliation{$^{\dag}$Department of Chemistry and Chemical Engineering, Royal Military College of Canada, 11 General Crerar Crescent, Kingston, ON, Canada, K7K~7B4\\
$^{\ddag}$Department of Mechanical Engineering, McGill University,	817 Sherbrooke St. West, Montreal, QC, Canada, H3A 0C3\\
$^{*}$Department of Mechanical and Materials Engineering, Queen’s University, 130 Stuart Street, Kingston, ON, Canada, K7P~2M4}

\usepackage{fancyhdr}
\begin{document}
\twocolumn[
\setlength{\fboxrule}{0.5pt}
\begin{@twocolumnfalse}
\maketitle
\begin{center}
\parbox{5in}{\textbf{Abstract. }%
The potential for reactive metal fuels to enhance the energetic output of high explosives has generated an enduring interest in the study of composite explosives. It has typically been demonstrated that added metal fuels can have little or even deleterious impact on the accelerating ability of composite military explosives relative to baseline performance. Often this has led to the assumption of limited reaction of the metal fuel over microsecond timescales. The widespread availability of Photonic Doppler Velocimetry has enabled time resolved measurement of accelerated confinement, ultimately demonstrating prompt reaction of metal fuels. Motivated by this observation, hydrocode modelling studies, and prior author's modifications of Taylor's tubular bomb model, we developed a differential equation form of Taylor's model in a manner where it is straightforward to add sources or phases. An afterburning version of the JWL equation of state was used to add energy to the gaseous products at a linear, time-dependent rate. The metal particles are assumed to remain in velocity equilibrium with the gaseous products and do not transfer heat or influence chemical composition. We focus exclusively on added aluminum as it remains the most ubiquitous choice of metal fuel. The model is initialized with a CJ state calculated from Cheetah 2.0 assuming the Al particles are inert in the detonation. JWL coefficients for the baseline explosive are also used. Qualitative agreement is observed between the model and previously published experiments.
% An empty line needed before the sectionline command

\sectionline}
\end{center}
\end{@twocolumnfalse}
]

%\printlength{\textwidth}\\
%\printlength{\columnwidth}

\thispagestyle{fancy}
\section{Introduction}
\label{Sec:TaylorIDS_Intro}

Adding reactive metal fuel, typically atomised aluminum, to high explosives (HE) to increase energetic output is well-established and extensively studied. However, mesoscale mechanisms and reaction kinetics for metal fuels remain unresolved. Assumptions about these mechanisms influence estimates for total energy release and where in the detonation product expansion it occurs. The fundamental observation that added aluminum can react sufficiently quickly to influence acceleration ability but not necessarily increase performance over pure explosives was made by Finger et al.\cite{Finger1970}.
\clearpage
\pagestyle{plain}
However, the influence of anaerobic metal fuel reaction on the accelerating ability of a composite HE remains controversial\cite{ErmolaevSurprise2000}. While metals oxidized by \ce{H2O}, \ce{CO2}, and \ce{CO} release more specific energy than detonating HE, adding metal:  1) reduces product gasses available upon detonation by dilution; 2) may reduce product gases through molar decrementing reactions; 3) deposits energy over a longer timescale while pressure accelerating the confining wall drops rapidly; 4) releases energy that may be trapped as latent heat in a solid product; and 5) diverts momentum from the products to particle drag. Thus, while added metal may react promptly, cancelling effects can limit performance.

The widespread availability of photonic Doppler velocimetry (PDV) has enabled time-resolved measurements of confiner acceleration with simplified experimental setup. Studies using PDV have conclusively demonstrated metal fuel reactions over microsecond timescales, albeit indirectly, through comparisons to velocimetry traces obtained with the simple explosive or an inertly diluted composite\cite{Manner2012,Tappan2016,LoiseauDetSymp2018_NM_AL,ZhangPartSize2024}. Rapid reaction is also supported by measurement of detonation product electrical conductivity\cite{Gilev2002,Gilev2006}. Some studies also suggest that a portion of the fuel may react within the chemical reaction zone of the detonation based on anomalous changes in detonation velocity\cite{DavydovTheo19992,Cowperthwaite1993,Gonthier2006}. This may be attributed to a reduction in gaseous product as solid oxides form\cite{Tappan2016}. Experimental studies have shown a weak effect of particle size on metal fuel energetic output, further confounding potential reaction models\cite{LoiseauDetSymp2018_NM_AL,ZhangPartSize2024,PeiPartsize2020}. Surface oxide passivation likely also plays a role in fuel reactivity\cite{LewisNano2013}.

Numerous modelling methodologies have attempted to resolve these complexities. Thermochemical equilibrium calculations (e.g. Cheetah\cite{Tappan2016}, EXPLO5\cite{SuceskaEXPLO2021}) can reasonably estimate detonation parameters and product expansion behaviour for some composite HEs. Semi-analytic techniques can also estimate metal fuel involvement in the detonation process\cite{Gonthier2006,Cowperthwaite1983}. Detailed multiphase calculations are likely necessary to resolve the non-equilibrium effects of adding large solid fractions to HE\cite{Mader1983,Mader2007,Ripley2012}. Multiphase hydrocode simulations that include transport between the particles and gaseous products and employ program-burn energy release by the particles have shown good agreement with experiment. Successive refinements of these two-phase models and different energy release assumptions have resulted in varying predictions for the involvement of Al reaction depending on mass fraction of metal particles and the simple HE studied, e.g.\,: nitromethane thickened with polyethylene glycol and mixed with 10-\textmu{m}-dia Al in the cylinder test\cite{MilneDet2002}; HMX/wax/50-\textmu{m}-dia Al in the cylinder test\cite{MilneDet2010}; nitromethane gelled with poly(methyl methacrylate) and mixed with 50-\textmu{m}-dia Al in slab tests\cite{PontalierEdenSlab2020}. The latter studies suggest 60\textendash{70\%} of the Al at 15\% mass loading and $\sim 25\%$ of the Al at 30\% mass loading reacts on wall acceleration timescales.

In the present study, we have attempted to capture some of the qualitative behaviour of rapid metal particle reaction in detonation products with a simple semi-analytic model. A permutation of Taylor's tubular bomb model\cite{TaylorTubular1963} was combined with the Zeldovich-von Neumann-D\"{o}ring (ZND) equations to treat the detonation product flow. This method allows source terms to be added easily. An afterburning version of the Jones-Wilkins-Lee (JWL) equation of state was used to treat the detonation products with energy added using programmed burn. Taylor's method is of historical interest and simple models may be useful for quick initial fitting of EOS coefficients before refinement using computationally expensive hydrocode iterations\cite{JacksonAnalytic2015}.

\section{Taylor Theory}

\begin{figure*}[t!]
	\centering
	\includegraphics[width=0.75\linewidth]{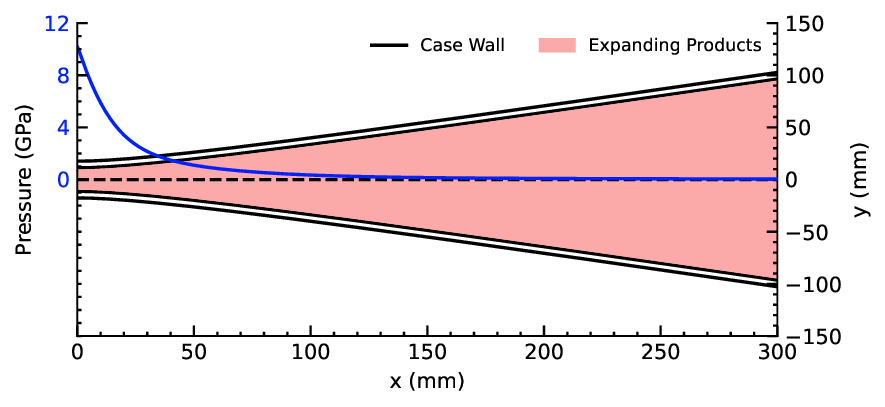}
	\caption{Typical wall shape for a pair of 6.35-mm-thick aluminum flyer plates accelerated by a 22.5-mm-thick slab of gelled nitromethane. Note the absence of wall thinning in the slab geometry. Also plotted is the spatial variation in detonation product pressure throughout the expansion process.}
	\label{fig:caseshapefulltextwidth}
\end{figure*}

Taylor developed a quasi-1D model to calculate the motion of a cylindrical wall accelerated by an axially detonating HE\cite{TaylorTubular1963}; simultaneously recognizing the diagnostic value of the geometry for measuring detonation product expansion ahead of widespread acceptance of the cylinder test (CYLEX)\cite{Kury1965,Lee1968}. Taylor's model has seen relatively little direct use: Allison et al. compared the model to explosively driven cylinders of varying wall thicknesses, treating the products as a polytropic gas\cite{Allison1960I,Allison1960II}. Baker et al. developed permutations of Taylor's model for various axisymmetric geometries, while incorporating more realistic equations of state\cite{Baker1993,Baker2010,Baker2010theory}. Baker also extended Taylor's model to asymmetric geometries with a Lagrangian plane of zero gas velocity\cite{Baker1993}. Taylor's kinematic relationships for wall tilt angle, and lateral and longitudinal velocity components are foundational for reconstructing wall trajectories in experiments instrumented with PDV\cite{JacksonAnalytic2015,ChaosKinematics2022}, and analytic predictions of warhead behaviour\cite{Walters1968,WaltersFCS1989}.

Taylor postulated that in a detonation-fixed frame of reference, the confining wall exits the detonation plane at the detonation velocity, $D$, with a small angle of tilt, $\theta$, where the lateral velocity component of the wall is small compared to $D$. Curvature of the wall is caused by a centripetal force generated by the pressure of the detonation products acting on an internal differential wetted area. The pressure along the length of the expanding confiner is governed by the strong form of the Bernoulli equation, coupled with the integral of the chosen pressure vs. specific volume relationship, determined from the product EOS. A typical casing shape from the present Taylor model is shown in Figure~\ref{fig:caseshapefulltextwidth}. 

Incorporating time-dependent reactions into this method is challenging since an integro-differential equation would result. Instead we have opted to treat the detonation product flow with the ZND equations for conservation of mass, momentum, and energy. With quasi-one dimensional area change to account for confiner expansion, these equations are:

\begin{align}
	\pde{}{x}(\rho u) =& -\rho u \frac{1}{A} \pde{A}{x}	
	\\
	\pde{}{x}(\rho u u + p) =& -\rho u u \frac{1}{A} \pde{A}{x}
	\\
	\pde{}{x}(\rho Y_j u) =& \, \rho \dot{\lambda_\mathrm{j}} - \rho u Y_{\mathrm{j}} \frac{1}{A} \pde{A}{x}	
\end{align}

\begin{align}
	\begin{split}
	\pde{}{x}\left( \rho u \left(e+\frac{1}{2} u^2 + \frac{p}{\rho}\right) \right) = \\ 
	-\rho u \left(e+\frac{1}{2}u^2 + \frac{p}{\rho} \right) \frac{1}{A} \pde{A}{x}	
	\end{split}
\end{align}	
with density $\rho$, velocity $u$, pressure $p$, cross-sectional area $A$, specific internal energy $e$, mass fraction $Y$ of species $j$, and reaction rate $\dot{\lambda_\mathrm{j}}$. Partial derivatives are taken relative to the position behind the detonation in the detonation-fixed frame.

Useful manipulations of the mass, momentum, and energy equations are:
\begin{align}
	\pde{\rho}{x} =& - \frac{\rho}{u} \pde{u}{x} - \rho \frac{1}{A} \pde{A}{x}
	\label{Eq:SimpMass} ,
	\\
	\pde{p}{x} =& - \rho u \pde{u}{x} ,\mathrm{\ and}
	\label{Eq:SimpMV}
	\\
	u \pde{u}{x} =& - \pde{e}{x} - \pde{}{x}\left(\frac{p}{\rho}\right).
	\label{Eq:SimpE}
\end{align}

\subsection{Equation of State}

The JWL equation of state was chosen given its simplicity and ubiquity. The authors found independent reviews of the JWL EOS by Weseloh\cite{WeselohJWL2014}, Mennikoff\cite{MenikoffJWL2017}, Segletes\cite{SegletesJWL2018}, and Farag\cite{FaragJWL2024} et al. helpful when manipulating the equation. In compact notation the $p(e,v)$ form can be written as:
\begin{equation*}
	p = \sum_{i}^{2} \left\{ \Lambda_i \left(1 - \frac{\omega}{R_i} \frac{\rho}{\rho_0}\right) e^{-\left(R_i \frac{\rho_0}{\rho}\right)}\right\} +\omega \rho (e-e_0)
	\label{Eq:JWL}
\end{equation*}
where $\rho_0$ is the undetonated explosive density, $\omega$ is the constant Gr\"{u}neisen parameter,  $\Lambda_1$, $\Lambda_2$, $R_1$, $R_2$ are fitted constants, and $e_0\approx{-q}$, where $q$ is the heat of reaction. In the present study, metal particle reaction is assumed to add energy through the $e_0$ term such that its derivative is non-zero. 

The partial derivative of the detonation product energy with respect to location behind the detonation can thus be calculated as:
\begin{equation*}\begin{split}
\pde{e}{x} =& \pde{}{x} \biggl[ e_0 + \frac{1}{\omega}\frac{p}{\rho} 
\\
+&\sum_{i}^{2} \left\{ \Lambda_i \left(\frac{1}{R_i \rho_0}-\frac{1}{\omega\rho}\right) e^{-\left(R_i\frac{\rho_0}{\rho}\right)}\right\}\biggr]
\end{split}\end{equation*}

Symbolic differentiation yields the following PDE for the product energy:
\begin{align}\begin{split}
	\pde{e}{x}&= \pde{e_0}{x} + \frac{1}{\omega} \pde{}{x}\left(\frac{p}{\rho}\right) 
	\\	
	+&\sum_{i}^{2} \left\{ \Lambda_i \left( \frac{\omega+1}{\omega\rho^2} - R_i\frac{\rho_0}{\omega\rho^3}\right) e^{-\left(R_i \frac{\rho_0}{\rho}\right)}\right\}\pde{\rho}{x} 
	\\
	&= \pde{e_0}{x} + \frac{1}{\omega}  \pde{}{x}\left(\frac{p}{\rho}\right) - \frac{1}{\omega} \frac{B(\rho)}{\rho^2} \pde{\rho}{x}
\end{split}\label{Eq:dedx}\end{align}
where the following placeholder function is used:
\begin{equation*}
	B(\rho) = \sum_{i}^{2} \left\{\Lambda_i \left(R_i \frac{\rho_0}{\rho} - (\omega+1)\right) e^{-\left(R_i \frac{\rho_0}{\rho}\right)} \right\}
\end{equation*}
Substitution of Equation~\ref{Eq:dedx} into Equation~\ref{Eq:SimpE} yields:
\begin{equation*}
	u \pde{u}{x} = - \pde{e_0}{x} + \frac{1}{\omega} \frac{B(\rho)}{\rho^2} \pde{\rho}{x} - \frac{\omega + 1}{\omega}\pde{}{x}\left(\frac{p}{\rho}\right)
\end{equation*}
Product-rule expansion of the derivative of the work term yields:
\begin{align*} 
	\pde{}{x}\left(\frac{p}{\rho}\right) = \frac{1}{\rho} \pde{p}{x} + p \pde{}{x}\left(\frac{1}{\rho}\right)
	= \frac{1}{\rho} \pde{p}{x} - \frac{p}{\rho^2} \pde{\rho}{x}
\end{align*}
such that:
\begin{align*}
	\begin{split}
	u \pde{u}{x} =& - \pde{e_0}{x} + \frac{1}{\omega} \frac{B(\rho)}{\rho^2} \pde{\rho}{x} \\&- \frac{\omega + 1}{\omega}\left(\frac{1}{\rho} \pde{p}{x} - \frac{p}{\rho^2} \pde{\rho}{x}\right)
	\end{split}
	\\
	\quad  = -\pde{e_0}{x} + &\frac{B(\rho) + p(\omega + 1) }{\omega \rho^2} \pde{\rho}{x} - \frac{(\omega + 1)}{\rho \omega} \pde{p}{x}
\end{align*}
The density (\ref{Eq:SimpMass}), and pressure (\ref{Eq:SimpMV}) differentials are then substituted to yield:
\begin{equation}
\begin{split}
	\pde{u}{x} = \left(\omega \pde{e_0}{x} + \frac{B(\rho) + p(\omega + 1)}{\rho} \frac{1}{A} \pde{A}{x}\right)& \bigg/ \\ \left(u - \frac{B(\rho) + p(\omega + 1)}{\rho u}\right)&
\end{split}
\end{equation}

\subsection{Particle reaction}
%\begin{equation}
%	\omega\rho(e+\lambda{q}); \quad  \frac{d\lambda}{dt} =a (1-\lambda)^{1/2} p^{1/6}
%\end{equation}
The ZND equations are typically solved by including a chemical kinetics mechanism for evolving gaseous product species\cite{Cowperthwaite1983}. By using this method, source terms such as mass, momentum, and energy transfer between phases can easily be added following the body of work using multiphase ZND equations. Presently, changes in the gaseous species are ignored and energy from particle reaction is added at a constant rate to the gasses with no change in particle mass nor exchange of momentum or heat between the particle and gas phases. This is similar to the Miller\cite{MillerJWLAB1992,Meaney2024} extension for afterburning. A simple linear burn model was assumed:
\begin{equation}
	\mde{Y_{\mathrm{m}}}{t} = \dot{\lambda}_{\mathrm{m}} = \frac{1}{\tau_{\mathrm{b}}}
\end{equation}
Thus, after converting from a time to spatial derivative, the energy release is defined as:
\begin{equation}
	\pde{e_0}{x} =\phi q_\mathrm{m} \frac{1}{u}{\mde{Y_{\mathrm{m}}}{t}}= \phi q_\mathrm{m} \frac{1}{u}\frac{1}{\tau_\mathrm{b}}
\end{equation}
where $\phi$ is the initial mass fraction of metal, $q_\mathrm{m}$ is the specific energy release of Al (set to 10.6 KJ/g\cite{PontalierEdenSlab2020} in this study), and $\tau_\mathrm{b}$ is the burn time, ranging from 25\textendash{100~\textmu{s}}.

This analysis implies that the metal particles are assumed to remain in velocity equilibrium with the gaseous product and do not depend on heat transfer to deliver energy to the product gases. Influence of particle oxidation on the composition and total moles of gaseous products is also neglected. Formation of new condensed products from particle reaction is also not considered.

\subsection{Cross-sectional area}

\begin{figure}
	\centering
	\includegraphics[width=1.0\columnwidth]{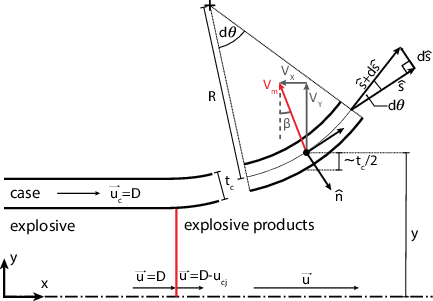}
	\caption{Geometry and flow parameters for the Taylor model. The lab-frame total metal velocity ($V_\mathrm{m}$) and the lateral ($V_\mathrm{Y}$) and longitudinal ($V_\mathrm{X}$) velocity components are included.}
	\label{fig:taylorflow}
\end{figure}

Motion of the wall is treated using an analogy to Taylor's equation of motion, but solved in stream-tube coordinates, and as a function of cross sectional area rather than the casing deflection angle, $\theta$, since this is required for coupling with the ZND equations. Only the sandwich geometry is presented in this analysis, where two walls (flyer plates) are accelerated by a slab of HE. It is assumed the walls diverge away from the centerline with a growing rectangular cross-sectional area, $A$, between them. In this geometry no wall-thinning occurs. The present analysis could also be extended to the more common cylindrical geometry by following the same derivation but where wall thinning must be included. 

Referring to Figure~\ref{fig:taylorflow}, consider a streamline through the center of the wall thickness as it expands away from the charge centerline. We adopt most of the assumptions from Taylor's original model: incompressible wall, small deflection angle, and small lateral velocity relative to the detonation velocity because of the quasi-1D assumption in ZND. In stream tube coordinates the wall only flows in the streamline direction ($\hat{s}$), greatly simplify the momentum equation for the wall, such that:
\begin{align} 
	\frac{1}{2} \pde{u_{\mathrm{c}}^2}{s} = - \frac{1}{\rho_{\mathrm{c}}} \pde{p}{s}
	&& \mathrm{and} &&
	\frac{u_{\mathrm{c}}^2}{R} = - \frac{1}{\rho_{\mathrm{c}}} \pde{p}{n}
	\label{eq:tubediffs} 
\end{align}
where $u_\mathrm{c}$ is the wall velocity, $R$ is the instantaneous radius of curvature of the wall, and $\rho_\mathrm{c}$ is the density of the wall meterial. The streamline component is Bernoulli's equation and the normal component is the centrifugal force from the pressure gradient across the case thickness. The latter is equivalent to Taylor's original equation of motion. Following Dehn~\cite{Dehn1984,DehnDetSymp1985}, $R$ can be written exactly through:
\begin{equation}
	\frac{1}{R} = \left| \left(\ode{s}{\theta}\right)^{-1} \right|
	= \frac{y''}{(1+y'^2)^{\frac{3}{2}}}
	\label{eq:curvature}
	%= \frac{\ode{}{x}(\ode{r}{x})}{(1+(\ode{r}{x})^2)^{\frac{3}{2}}}.
\end{equation}

Since the coordinates are relative to the centreline of the wall, the expanding detonation product thickness is approximately $2(y - \frac{t_{\mathrm{c}}}{2})$ so long as $\theta$ is reasonably small. Neglecting product expansion off the edges of the HE slab, the instantaneous detonation product thickness can be related to the cross sectional area by:
\begin{equation}
	A = w_{\mathrm{c}} \left(y - \frac{t_{\mathrm{c}}}{2}\right)
\end{equation}
where $w_{\mathrm{c}}$ is the width of the slab. Assuming constant wall thickness $t_{\mathrm{c}}$, the derivatives:
\begin{align*}
	y' =& \ode{}{x} \left(\frac{A}{w_{\mathrm{c}}} + \frac{t_{\mathrm{c}}}{2}\right) = \frac{A'}{w_{\mathrm{c}}}
	\\
	y'' =& \ode{}{x} \left(\frac{A'}{w_{\mathrm{c}}}\right) = \frac{A''}{w_{\mathrm{c}}}	
\end{align*}
are substituted into Equation~\ref{eq:curvature}, yielding:
\begin{equation}
	\frac{1}{R} = \frac{\frac{A''}{w_{\mathrm{c}}}}{\left(1+\left(\frac{A'}{w_{\mathrm{c}}}\right)^2\right)^{\frac{3}{2}}}
	=  \frac{w_{\mathrm{c}}^2 A''}{\left(w_{\mathrm{c}}^2 + A'^2\right)^{\frac{3}{2}}}
	\label{eq:dA}
\end{equation}
Equation~\ref{eq:dA} can be combined with Equation~\ref{eq:tubediffs} to relate the pressure on the wall to the cross-sectional area and its derivatives by integrating through the wall thickness, where curvature $1/R$, wall velocity $u_{\mathrm{c}}$, and wall density $\rho_{\mathrm{c}}$, are uniform. Noting the reversed signs on the integration bounds since $\hat{n}$ is positive away from the centre of curvature:
\begin{align}
	\int_p^{p_{\mathrm{s}}}\partial p =& \int_{{y+\frac{t_{\mathrm{c}}}{2}}}^{{y-\frac{t_{\mathrm{c}}}{2}}} - \rho_{\mathrm{c}} \frac{u_{\mathrm{c}}^2}{R} \partial n
	\\
	p - p_{0}=& \rho_{\mathrm{c}} t_{\mathrm{c}} \frac{u_{\mathrm{c}}^2}{R}
	\label{eq:intnorm}
\end{align}
where $\rho_{\mathrm{c}} t_{\mathrm{c}}$ is the wall mass per unit area, and $p_0$ is the surrounding pressure, which we assumed to be zero but retained in subsequent equations.

\begin{figure*}[t!]
	%\begin{strip}
	\begin{align*}
		\hspace{0.2\textwidth}
		%\ode{u}{x} =& \omega(u - \frac{B + (\omega + 1) p}{\rho u} )^{-1} (\frac{1}{u} \sum_j (e_{0,j} \mde{Y_j}{t})) + \frac{1}{\rho} \frac{ (B + (\omega + 1) p)}{(u - \frac{B + (\omega + 1) p}{\rho u} )} \frac{A'}{A}
		\ode{u}{x} =& \left(\omega \pde{e_0}{x} + \frac{B(\rho) + p(\omega + 1)}{\rho} \frac{1}{A} \ode{A}{x}\right)\bigg/ \left(u - \frac{B(\rho) + p (\omega + 1) }{\rho u} \right)
		\\
		\ode{e_0}{x} =& \, \phi q_\mathrm{m} \frac{1}{u}\frac{1}{\tau_\mathrm{b}}
		\\
		\ode{p}{x} =& - \rho u \ode{u}{x}
		\\
		\ode{\rho}{x} =& - \frac{\rho}{u} \ode{u}{x} - \rho \frac{A'}{A}
		\\
		\ode{A'}{x} =& \, A'' = \left(\frac{p - p_{\mathrm{0}}}{\rho_{\mathrm{c}} t_{\mathrm{c}}}\right)  \left(\frac{(w_{\mathrm{c}}^2 + A'^2)^{\frac{3}{2}}}{ w_{\mathrm{c}}^2}\right) \bigg/ \left(D^2  + \frac{2}{\rho_{\mathrm{c}}}(\kappa p_{\mathrm{CJ}} - p)\right)
		\\
		\ode{A}{x} =& \, A'
	\end{align*}
	\hspace{0.2\textwidth}Recall:
	\begin{equation*}
		\hspace{0.2\textwidth}
		B(\rho) = \sum_{i}^{2} \left\{\Lambda_i \left(R_i \frac{\rho_0}{\rho} - (\omega+1)\right) e^{-\left(R_i \frac{\rho_0}{\rho}\right)} \right\}
	\end{equation*}
	\hspace{0.2\textwidth}With initial conditions: 	
	\begin{equation*}
		\hspace{0.2\textwidth}
		u = (D-u_{\mathrm{CJ}}); \quad	p = p_{\mathrm{CJ}};\quad \rho = \rho_{\mathrm{CJ}};\quad A' = 0;\quad A = A_{\mathrm{charge}}
	\end{equation*}
	%\end{strip}
	\caption{Summary of the system of equations and initial conditions used in the present model.}
	\label{fig:Eqs}
\end{figure*}

The wall velocity can then be related to the pressure and curvature by integrating Bernoulli's equation along the stream-tube coordinate, again assuming the case is incompressible and the velocity through the thickness is uniform:
\begin{align}
	\int_{D^2}^{v_{\mathrm{c}}^2} \partial u_{\mathrm{c}}^2 =& - \int_{ \kappa p_{\mathrm{CJ}}}^p \frac{2}{\rho_{\mathrm{c}}} \partial p
	\\
	u_{\mathrm{c}}^2 =& D^2  + \frac{2}{\rho_{\mathrm{c}}}(\kappa p_{\mathrm{CJ}} - p)
	\label{eq:intstream}
\end{align}
while taking the state immediately behind the detonation as the lower integration bound, where ${u_{\mathrm{c}} = D}$ and $p = p_{\mathrm{CJ}}$. In reality, early confiner motion is compressible because of the reverberating shock transmitted by the detonation. Pressure at the product\textendash{wall} interface is governed by shock impedances and obliqueness of the transmitted shock, but is typically some fraction of the detonation pressure\cite{NealDetSymp1976}. To account for these experimental realities we introduced a fitting constant, $\kappa$. In comparison to experiments $\kappa = 1.9$ gave best agreement; a value much higher than would be expected from results from Neal~\cite{NealDetSymp1976}. We have not resolved this inconsistency, nor examined values for other experimental explosive-metal pairs. Note that if the pressure term in Equation~\ref{eq:intstream} is neglected, Taylor's original assumption, ${u_{\mathrm{c}}=D}$,  is recovered.

Equations~\ref{eq:dA},\ref{eq:intnorm}, and \ref{eq:intstream} can be combined:
\begin{align*}
	p - p_{\mathrm{0}} =& \frac{\rho_{\mathrm{c}} t_{\mathrm{c}}}{R} \left(D^2  + \frac{2}{\rho_{\mathrm{c}}}(\kappa p_{\mathrm{CJ}} - p)\right)
\end{align*}
and the second derivative of area isolated to yield a final closure equation as a function of detonation product pressure:
\begin{equation}
	\begin{split} 
	A'' = \left(\frac{p - p_{\mathrm{0}}}{\rho_{\mathrm{c}} t_{\mathrm{c}}}\right) & \left(\frac{(w_{\mathrm{c}}^2 + A'^2)^{\frac{3}{2}}}{ w_{\mathrm{c}}^2}\right) \bigg/ \\ & \left(D^2  + \frac{2}{\rho_{\mathrm{c}}}(\kappa p_{\mathrm{CJ}} - p)\right)
	\end{split}
\end{equation}

For convenience, the complete system of equations and initial conditions is summarized in Figure~\ref{fig:Eqs}. In the current study the system of equations was solved numerically in Python using the Radau method from the SciPy package ODE solver.

\section{Comparison to Experiment}
\label{Sec:TaylorIDS_ComparisontoExp}
\begin{figure}
	\centering
	\includegraphics[height=155.4965pt]{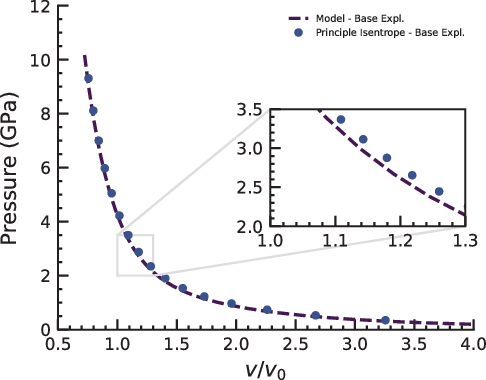}
	\caption{Principle isentrope for the baseline explosive (dots), plotted against the p($v$) history obtained from the model (dashed line).}
	\label{fig:psvsmodel}
\end{figure}

Confiner wall velocity is now typically measured using PDV. For a probe observing at 90\textdegree~from the initial position of the wall, only the radial velocity component in the lab frame ($V_\mathrm{Y}$ in Fig.\ref{fig:taylorflow}) is measured directly. This velocity is equivalent to the derivative of $y$ in the present model's detonation fixed frame. The detonation-fixed frame spatial derivative can be related to the lab-fixed time derivative via:
\begin{align*}
	\left. \pde{}{t}\right|_{\mathrm{lab}} = D \pde{}{x}
\end{align*}
such that the lateral velocity from the model can be compared to the PDV velocity history for the sandwich geometry using:
\begin{equation}
	\left. V_{\mathrm{c},Y} \right|_{\mathrm{lab}} = D \ode{}{x} \left(y + \frac{t_{\mathrm{c}} }{2}\right) = D \frac{A'}{w_{\mathrm{c}}}
\end{equation}
We compare the present model with symmetric sandwich experiments from Loiseau et al.\cite{LoiseauDetSymp2018_NM_AL}. The subset of experiments presently considered used 6.35-mm-thick 6061 aluminum flyer plates. The test explosive was nitromethane (NM) gelled with 4\% poly(methyl methacrylate) by mass. The gelled NM was sensitized with 0.5\% 3M K1 glass microballoons (GMB) by mass and 15\%, or 30\%  Valimet H-50 aluminum powder was added by mass to the sensitized mixture. An inert control consisting of 20.5\% alumina powder by mass is also considered. The explosive cavity was initially 22.5\,mm thick and the sandwich had a width of 10.2\,cm. The PDV probes measured at 90\textdegree.

Cheetah 2.0 was used to determine JWL coefficients for the \textit{baseline} explosive (0\% Al), alumina control, and 15\% or 30\% Al assuming either complete reaction or entirely inert behaviour for the Al. The GMB were treated as porosity. JWL coefficients are shown in Table~\ref{tbl:JWL}.

\begin{table*}[tb]
	\centering
	\begin{threeparttable}%
	\caption{Summary of detonation properties and JWL coefficients.}
	\label{tbl:JWL}
	\setlength{\tabcolsep}{2.5pt}
	\begin{tabular}{|c|c|c|c|c|c|c|c|c|c|c|c|}
			\hhline{|-|-|-|-|-|-|-|-|-|-|-|-|}\Ts
		Explosive & $\rho_0$ & $D$ & $P_\mathrm{cj}$ & $u_\mathrm{cj}$ & $\rho_\mathrm{cj}$ & $\Lambda_1$ & $\Lambda_2$ & $C$ & $R_1$ & $R_2$ & $\omega$ \\
		~& g/cc & km/s& GPa & km/s & g/cc & GPa & GPa & GPa &- &-&-\\
			\hhline{|=|=|=|=|=|=|=|=|=|=|=|=|}
		Baseline &1.09 & 5.80 & 10.17 & 1.60 & 1.51 & 190.48 & 3.93 & 1.08 & 4.58 &1.04 & 0.35\Ts\\
		20.5\% Alumina & 1.29 & 5.20 & 8.92 & 1.33 & 1.73 & 248.20 & 3.56 & 0.91 & 4.94 & 1.08 & 0.28\\
		15\% Al inert & 1.20 & 5.42 & 8.96 & 1.38 & 1.61 & 293.37 & 4.75 & 0.94 & 5.25 & 1.17& 0.30\\
		30\% Al inert & 1.33 & 5.13 & 7.91 & 1.16 & 1.72 & 293.86 & 1.59 & 0.45 & 4.84 & 0.64 & 0.12\\
		15\% Al active &  1.20 & 5.91 & 12.03 & 1.70 & 1.68 & 159.19 & 2.18 & 1.33 & 3.96 & 0.72 & 0.27\\		
			\hhline{|-|-|-|-|-|-|-|-|-|-|-|-|}
	\end{tabular}
	\end{threeparttable}
\end{table*}

An initial validation was performed to confirm that the present model reproduces expansion along the principle isentrope when no afterburning energy is added. For the baseline explosive, the principle isentrope was calculated via:
\begin{equation}
	p_\mathrm{s} = \Lambda_1 e^{-R_1 \frac{\rho_0}{\rho}} + \Lambda_2 e^{-R_2 \frac{\rho_0}{\rho}} + C \left({\frac{\rho_0}{\rho}}\right)^{-(1+\omega)} 
	\label{eq:JWLpi}
\end{equation}
and is plotted against the $p(v)$ history from the model in Figure~\ref{fig:psvsmodel}. Agreement is reasonable over the accessible range of relative volumes.

\begin{figure}[t!]
	\centering
	\includegraphics[height=155.4965pt]{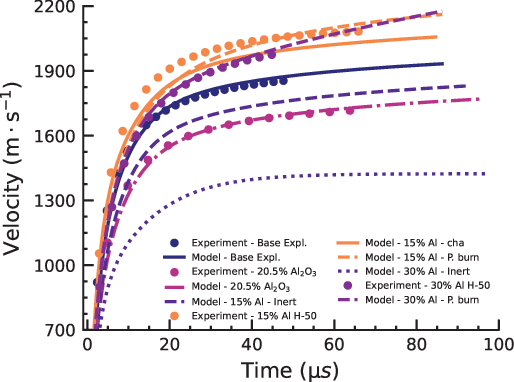}
	\caption{Experimental velocity histories (dots) plotted against the predictions of the present model.}
	\label{fig:Veltraces}
\end{figure}

Figure~\ref{fig:Veltraces} shows model predictions for wall acceleration plotted as lines versus experimental velocity histories plotted as dots. Model predictions using Cheetah to treat Al reaction are denoted by ``-cha'', those using programmed burn are donated by ``-P. burn''. Note the smooth ballistic acceleration of the wall in these experiments. This is because of the relatively low brisance of the NM explosive and the detonation being approximately sonic relative to the sound speed in aluminum. The characteristic surface oscillations from shock reverberation are thus suppressed. The model accurately predicts the acceleration history of the wall for the baseline case and also the inert control diluted with 20.5\% alumina. The accuracy for the inert control is surprising given the presumed importance of momentum and heat transfer during acceleration of the particles in the detonation and during product expansion.

\begin{figure}
	\centering
	\includegraphics[height=155.4965pt]{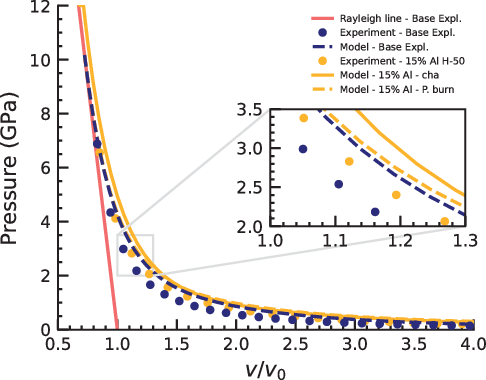}
	\caption{Comparison of $p(v)$ histories.\hfill}
	\label{fig:Isentropes}
\end{figure}

For 15\% Al reacting in equilibrium, model results using Cheetah-derived JWL coefficients agree reasonably well with the experimental results, but under-predict wall velocity after $\approx$ 10\,\textmu{s}. For the afterburning model predictions for 15\% Al, the baseline JWL parameters were combined with a burn time ($\tau_\mathrm{b}$) of 25\,\textmu{s}. The model again underpredicts the experimental result at early times but matches and then slightly exceeds the experimental values at probe cut-out. This suggests that the linear burn model adds too much energy, but also adds it too late in the expansion process. The influence of particle energy release on effective product pressure is shown in Figure~\ref{fig:Isentropes}, which plots the $p(v)$ histories of the model predictions, versus experimental isentropes extracted using the method outlined by Jackson\cite{JacksonAnalytic2015}. 

Cheetah yielded poor JWL fits for 30\% Al reacting in equilibrium, which resulted in non-physical predictions of wall velocity. These results are thus omitted from Figure~\ref{fig:Veltraces}. For the afterburning model predictions for 30\% Al, the baseline JWL parameters were again used, but a burn time of 80\,\textmu{s} was instead specified. Agreement was overall good, but continued acceleration of the wall after 50\,\textmu{s}, beyond the probe cut-off, is likely non physical.

\section{Concluding Remarks}

A simple, semi-analytic model was developed following the assumptions established by Taylor\cite{TaylorTubular1963}. When appropriate EOS coefficients were used, reasonable agreement with flyer plate experiments was observed. A simple linear burn model for Al particle reaction qualitatively suggests that a significant fraction of aluminum can burn to influence metal acceleration. 

\bibliographystyle{Det_Symp}
\bibliography{Taylor_IDS}

\noindent{\textbf{Question from Sorin Bastea, LLNL}}\\
Have you looked at the effect of Al particle size?
\\

\noindent{\textbf{Reply by Jason Loiseau}}\\
We have not yet investigated particle size in the model. The current implementation can only prescribe a burn time, and the correlation between burn time and particle size (e.g. extrapolations from Beckstead correlations to detonation products) is still controversial. We did investigate Al particle size effects experimentally and presented these results at the previous IDS: we saw a weak effect of particle size on the accelerating ability of the composite HE.
\\

\noindent{\textbf{Question from Tim Manship, Purdue University}}\\
Very fascinating approach! For your model, do you assume aluminum combustion is just adding energy to gas products or are you also accounting for addition of aluminum combustion products?
\\

\noindent{\textbf{Reply by Jason Loiseau}}\\
We made the simplifying assumption that Al combustion adds energy directly into the detonation product gases. We neglect any effects on the chemical composition of the detonation products, and neglect the formation and condensation of solid Al products or additional carbon as oxygen is scavenged from CO and CO\textsubscript{2}. We thus also neglected any heat transfer to/from particles or solid product.
\\

\noindent{\textbf{Question from Christopher Miller, LLNL}}\\
How well was the aluminum distributed throughout the samples and how would non-homogeneity influence your model?
\\

\noindent{\textbf{Reply by Jason Loiseau}}\\
Free-flowing powders mix well into the gel. Based on sample microscopy we have observed good uniformity and repeat trials have shown good reproducibility. Since the model is quasi-1D it cannot account for inhomogeneity along the height or width of the slab (or radially in cylinders). Longitudinal inhomogenity could be addressed by varying particle mass fraction and interpolating JWL parameters for different initial solid loadings.

\end{document}